# Spin imbalance of charge carriers induced by an electric current

Antonio Hernando,[a,b], F. Guinea[b,c] and Miguel A. García [d]

[a]Instituto de Magnetismo Aplicado, UCM-CSIC-ADIF, P. O. Box 155, 28230-Las Rozas. Madrid, Spain.
[b]IMDEA , Nanociencia, Faraday 9, 28049 Madrid, Spain.
[c]School of Physics and Astronomy, University of Manchester, Manchester, M13 9PY, UK.
[d]Instituto de Cerámica y Vidrio, CSIC, C/Kelsen 5, 28049-Madrid, Spain.

We analyze the contribution of the inhomogeneous magnetic field induced by an electrical current to the spin Hall effect in metals. The Zeeman coupling between the field and the electron spin leads to a spin dependent force, and to spin accumulation at the edges. We compare the effect of this relativistic correction to the electron dynamics to the features induced by the spin-orbit interaction. The effect of current induced magnetic fields on the spin Hall effect can be comparable to the extrinsic contribution from the spin-orbit interaction, although it does not require the presence of heavy elements with a strong spin-orbit interaction. The induced spins are oriented normal to the metal slab.

*Introduction.* The passage of an electric current in a metallic sheet leads to the accumulation of spin at its edges, the Spin Hall Effect (SHE). This phenomenon, related to the accumulation of charge when a current is induced in a ferromagnet, the Anomalous Hall Effect (AHE) has opened new ways of manipulating spins in metals. A natural explanation of the SHE and AHE can be given in terms of the spin-orbit interaction[1]. For the SHE, either spin dependent scattering due to impurities[2], or the action of an intrinsic interaction on accelerating carriers[3,4] can lead to the deflection of spin currents, and to spin accumulation at the edges, see Fig.[1]. The nature of the SHE and the AHE are discussed in detail in refs.[4,5].

In the following, we analyze an additional contribution to the SHE, arising from the existence of non-homogeneous magnetic fields within the metallic sheet, see Fig.[2]. These fields give rise to a spatially dependent Zeeman energy, and to lateral forces on the carriers, whose sign depends on the spin. As a result, a gradient in the spin density within the sample is induced. It will be shown that this mechanism gives rise to spin imbalances comparable to those derived from the extrinsic and intrinsic mechanisms mentioned earlier.

Our analysis does not rely explicitly on the existence of a strong spin orbit coupling, so that it can be generalized to metals made from light elements (note, however, that the Zeeman energy can be considered a relativistic effect which arises from the Dirac equation which describes the electrons). The main features of the model are described next. Then, semi quantitative estimates of the spin accumulation induced by inhomogeneous magnetic fields are made, and compared to the results derived from the contributions to the SHE from extrinsic and intrinsic spin-orbit interactions. Finally, the main conclusions and possible extensions will be presented. We use CGS units throughout the manuscript. This choice helps us to highlight the way in which relativistic corrections enter in the analysis.

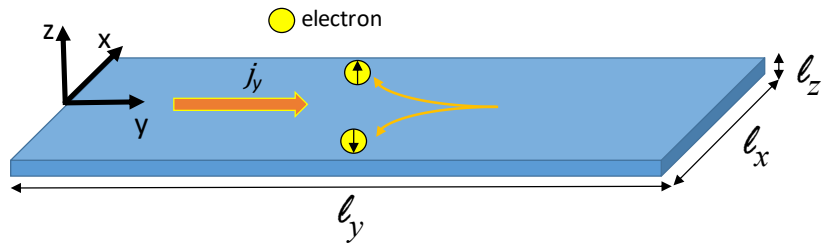

Figure 1. Sketch of the spin currents induced by the magnetic field created by an electrical current in a Hall bar.

*Inhomogeneous magnetic fields in current carrying metallic sheets.* We consider a thin metallic sheet, of dimensions $\ell_x$, $\ell_y$ and $\ell_z$, with $\ell_z \ll \ell_x, \ell_y$, see Fig.[1]. We assume a constant current density per unit area, $j_y$ along the $y$ direction. This current induces at the plane of the sheet, a magnetic field along the $z$ direction. In the limit $\ell_y \gg \ell_x, \ell_z$ and $\ell_z \ll \ell_x$, the field is[6] (see also the Supplementary Information)

$$B_z^I(x) = \frac{2 j_y \ell_z}{c} \log\left(\frac{\ell_x/2 + x}{\ell_x/2 - x}\right) \qquad (1)$$

The label $j$ stands for the current density, $j_y = I/(\ell_x \ell_z)$, where $I$ is the total current. Throughout most of the sheet the field can be well approximated by function proportional to the $x$ coordinate[6],

$$B_z^I(x) \approx 8 \frac{j_y x}{c} \frac{\ell_z}{\ell_x} \qquad (2)$$

This is the field measured in a frame of reference at rest. In the frame of reference where the current carriers are at rest a positive current, associated with the ions, can be defined. This current induces an opposite magnetic field, which is felt by the carriers. The Zeeman energy associated to this field is

$$E_Z^I = \int d^3\vec{r}\, \mu_B B_z(\vec{r})[n_\uparrow(\vec{r}) - n_\downarrow(\vec{r})] \quad (3)$$

Where $\mu_B = (e\hbar)/(2mc)$ is Bohr's magneton. As a result, a force acting on the spin of the carriers arises

$$F^I \approx \pm 8 \frac{\mu_B j_y}{c} \frac{\ell_z}{\ell_x} = \pm 8 \frac{\mu_B I}{c \ell_x^2} \quad (4)$$

This force is sketched in Fig.[2].

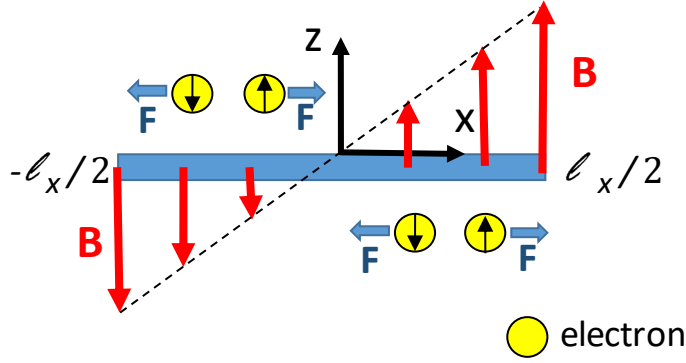

Figure 2. Sketch of the inhomogeneous magnetic field created by a current, and the associated Zeeman forces.

*Numerical estimates. Magnetic field.* The magnetic field induced by the current, given in eq.(1), is maximum at the edges of the sample, $x \approx \ell_x/2$,

$$B_{max} \approx 8 \frac{j_y \ell_z}{c} \quad (5)$$

For typical current densities, $j_y \approx 10^7$ A $\times$ m$^{-2}$ and $\ell_z \approx 100$ nm we obtain $B_{max} \approx 10^{-2}$ G. This field is much lower than typical fields used in experiments. We assume that the orbital polarization induced by a magnetic field of this magnitude is negligible. As mentioned before, we focus on the spin dependent forces induced by the *gradient* of this field.

*Numerical estimates. Hall angle.* The spin Hall angle, obtained from eq.(4), is described in the Supplementary Information. An approximation based on a three dimensional nearly free electron model for the carriers gives, see eq.(A5):

$$\alpha \approx \frac{8}{137^2} \frac{k_F a_B}{3\pi^2} k_F \ell \frac{\ell_z}{\ell_x} \quad (6)$$

Where $\alpha$ is the Hall angle, $k_F$ is the Fermi wavevector, $a_B \approx 0.53$ Å is the Bohr radius, and $\ell$ is the mean free path. The factor $137^2$ in the denominator of the r.h.s. in eq.(6) contains the fine structure constant, and it highlights the relativistic origin of the effect.

Eq.(6) applies to metallic samples with light elements, where only impurities with heavy elements can lead to a spin Hall effect. Experiments with Al samples were reported in[6]. We assume that $k_F^{Al} \approx 1.7$ Å$^{-1}$, $k_F^{Al} \ell \approx 10^2$ and $\ell_z/\ell_x \approx 1/5$. We obtain $\alpha \approx 2.3 \times 10^{-4}$. This value is of the same order of magnitude as the results reported in[8,9] (see also[10,11]). The estimate in eq.(6) is valid for nearly free electron systems, where the effective mass is approximately equal to the free electron mass, $m_{eff} \approx m_e$. The value of $\alpha$ in eq.(6) is enhanced by a factor $m_{eff}/m_e$ if the two masses differ significantly, as in transition metals.

A similar analysis to the one leading to eq.(6), but for a quasi-two dimensional electron gas where only one subband is occupied gives, see eq.(A6)

$$\alpha_{2D} \approx \frac{8}{137^2} \frac{1}{2\pi} k_F \ell \frac{a_B}{\ell_x} \quad (7)$$

The dependence of the angle on the fine structure constant implies that, $\alpha \propto c^{-2}$, where $c$ is the velocity of light. This is the same as the one expected for the extrinsic contribution from skew scattering effects, see below. This may complicate the experimental separation of the contribution from the field gradient discussed here from that of skew scattering. A plot of the dependence of $\alpha$ on elastic scattering and on the aspect ratio of the sample, using. Eq.(6), is shown in Fig.[3].

Ref.[2] suggested the measurement of the Spin Hall Effect through the voltage induced in a strip connecting the edges of the sample. This setup does not necessarily work for the detecting the contribution discussed here, as the current induced magnetic field can also affect the electrons in the strip, see discussion in the Supplementary Information.

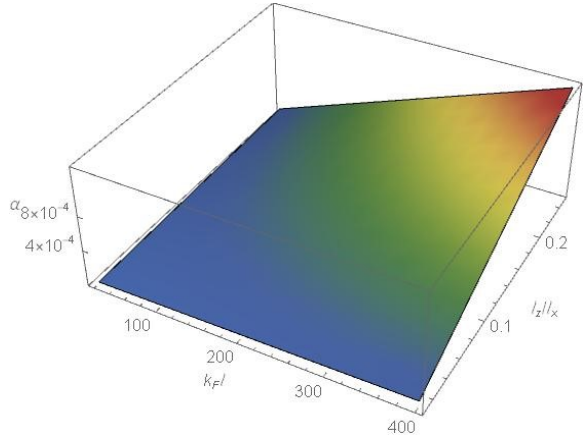

Figure 3. Spin Hall angle, $\alpha$, due to the current induced magnetic field, as function of the product of the Fermi wavelength and the mean free path, $k_F \ell$ and the aspect ratio of the sample, $\ell_z/\ell_x$. See equation (6).

*Numerical estimates. Spin accumulation at the edges.* The net spin at the edges, if the spin diffusion length is smaller than the width of the sample, $\ell_x$, is determined by the difference in the chemical potential for the spin up and spin down electrons, $\mu_\uparrow - \mu_\downarrow \approx 2\mu_B B_z \left(x \approx \pm \frac{\ell_x}{2}\right)$. The expression for the field is given in eq.(1). This expression has been obtained in the limit $\ell_z \ll \ell_x$ and it breaks close to the edges, when $\left|\frac{\ell_x}{2} - |x|\right| \leq \ell_z$. We take the average of $B_z$ over distances from the edge comparable to $\ell_z$, and obtain $\langle B_z \rangle \approx \frac{2j_y \ell_z}{c} \log\left(\frac{\ell_x}{\ell_z}\right)$. We finally obtain that the spin accumulation at the edges, in units of electrons per area is

$$n_\uparrow - n_\downarrow \approx \frac{4\mu_B j_y \ell_z}{c} \log\left(\frac{\ell_x}{\ell_z}\right) \times n(\epsilon_F) \quad (8)$$

Where $n(\epsilon_F)$ is the density of states of the metal at the Fermi energy. We assume $\ell_x = 1\mu$m, $\ell_z = 0.1\mu$m, $j_y = 10^3$ A/mm$^2$ and $n(\epsilon_F) = 1$eV$^{-1}$Å$^{-3}$. Inserting these parameters into eq.(8) we obtain $n_\uparrow - n_\downarrow \approx 10^{18}$cm$^{-3}$.

*Comparison to the skew scattering contribution to the Spin Hall Effect.* We first estimate the combined effect of the spin scattering interaction and defect scattering following ref.[2]. We assume that the effect of the impurities is roughly equivalent to the effect of a magnetic field equal to the magnetization due to the carriers with a given spin, $B_{eff} \approx \frac{e\hbar}{mc} n_\uparrow$ where $n_\uparrow$ is the density per unit volume of carriers with spin up. The associated force is

$$F_{ext}^{SH} \approx \frac{v e B_{eff}}{c} \approx \frac{v e^2 \hbar}{mc^2} n_\uparrow \quad (9)$$

Where v is the carrier velocity. Using $I \approx ev(n_\uparrow + n_\downarrow)\ell_x \ell_z$, we obtain

$$F_{extr}^{SH} \approx \pm \frac{\hbar e I}{mc^2 \ell_x \ell_z} \approx \pm \frac{\mu_B I}{c} \times \frac{1}{\ell_x \ell_z} \quad (10)$$

Which leads to a spin Hall angle

$$\alpha_{extr} \approx \frac{1}{137^2} \frac{k_F a_B}{3\pi^2} k_F \ell \quad (11)$$

So that, using eq.(6), we obtain

$$\frac{\alpha_I}{\alpha_{extr}} \approx 8 \frac{\ell_z}{\ell_x} \quad (12)$$

Note that this estimate of $F_{ext}^{SH}$ in eq.(10) gives an upper bound to its actual value, as the analysis assumes that the spin dependent scattering induced by impurities leads to an effective field comparable to the one generated by the total spin of the carriers in the metal. A general discussion of skew scattering effects, following the original ideas[12,13,14], can be found in[15]. A different estimate of $F_{ext}^{SH}$ based on a two dimensional nearly free electron model and the Boltzmann equation is given in the supplementary information, see eq.(A17), and it gives

$$\alpha_{extr}^{2D} \approx \frac{(k_F a_B)^2}{137^2} \frac{\ell}{\ell_{SOI}} \quad (13)$$

Where $\ell_{SOI}$ is a mean free path associated to the impurities which induce the skew scattering. Using this estimate, and comparing to eq.(7), we find

$$\frac{\alpha_I^{2D}}{\alpha_{extr}^{2D}} \approx \frac{8}{k_F a_B} \frac{\ell_{SOI}}{\ell_x} \quad (14)$$

It is finally interesting to note that three dimensional scattering by impurities with spin-orbit coupling, described generically by terms such as $V_{imp}(\vec{r}) \approx V_0 \vec{L} \vec{s} \delta(\vec{r})$, do not favor a particular spin orientation. This scattering is expected to dominate in metallic systems where many subbands are occupied, such that $\ell_z \gg k_F^{-1}$. In these samples, there is no preferred spin orientation which spontaneously will accumulate at the edges (note this in not the case for experiments where a spin polarized current is injected). On the other hand, the magnetic field induced by the current considered here selects a spin orientation, and always leads to a spin accumulation at the edges.

*Other contributions to the Spin Hall Effect.* A similar analytical semi-quantitative comparison between the current induced SHE discussed here and the intrinsic contribution of the spin orbit interaction is not possible. Simple estimates based on the Rashba interaction give vanishing effects[16], once leading effects due to impurities are taken into account[4,5,6]. More involved calculations using the Berry curvature in complex, spin dependent, models for the bands of transition metals suggest the the SHE conductivity can be written as[17,18]

$$\sigma_{xy}^z \approx \frac{e}{4a} \times \frac{\langle \vec{l} \vec{s} \rangle_{FS}}{\hbar^2} \quad (15)$$

Here, $a$ is the lattice constant, and $\langle \vec{l} \vec{s} \rangle_{FS}$ is the average over the Fermi surface of the product of the orbital angular momentum and the spin, which depends on the strength of the spin orbit interaction. The value of $\langle \vec{l} \vec{s} \rangle_{FS}$ is determined by the intrinsic spin orbit interaction in the material, which typically scales as $c^{-2}$. Hence, the current contribution to the spin Hall voltage, the contribution from skew scattering, and the intrinsic contribution all arise from relativistic effects, and show the same dependence on the velocity of light. Note that the intrinsic contribution changes the conductivity, see eq.(15).

There is finally a contribution to the spin Hall effect from side jump scattering. To our knowledge, there are not reliable techniques to extract simple estimates of the magnitude of these processes. We cannot compare the role of the current induced field considered here to that of side jump scattering.

*Conclusions.*

We have analyzed the contribution to the spin Hall effect of the magnetic field associated to the current flowing through a metallic sample. The spin accumulation at the edges arises from the Zeeman coupling to the inhomogeneous magnetic field which exists within the sample. The gradient of the Zeeman coupling acts like a force which accelerates the carriers in the direction perpendicular to the current, in opposite way for spin up and spin down electrons, where the spins are oriented along the magnetic field. This effect is relativistic in nature, and its contribution to the spin Hall angle depends on the value of the velocity of light as $\alpha \propto c^{-2}$, similarly to the (extrinsic) contribution from skew scattering processes.

The current induced spin accumulation is linearly proportional to the scattering time or the transport mean free path, also like the contribution from skew scattering. Hence, it should dominate in clean samples, with $k_F \ell \gg 1$. Estimates from simple models suggest that the effect of the current induced magnetic field should be at least comparable to that of skew scattering.

The current induced spin Hall effect depends on the aspect ratio of the sample, through the ratio $\ell_z/\ell_x$, where $\ell_z$ is the thickness, and $\ell_x$ is the width. Hence, it is not a unique property of the material. This can explain the broad distribution of Hall angles for the same material reported in the literature. In contrast with the effect of skew scattering, the current induced spin Hall effect does not require the existence of a strong spin-orbit interaction, either due to heavy ion impurities or intrinsic to the material. Hence, it should be present in any sample, and play a significant role in light metal materials, such as Al.

Samples which are far from being quasi-two dimensional, $k_F \ell_z \gg 1$, typically do not have a preferred spin orientation, so that transverse spin currents due to skew scattering will average to zero. The current induced magnetic field provides a preferred spin direction, enhancing the robustness of the spin Hall effect.

**Acknowledgements**

This work was supported by the Spanish Ministry of Innovation, Science and Technology and Spanish Ministry of Economy and Competitiveness through Research Projects MAT2015-67557-C2-1-P, MAT2017-86450-C4-1-R, S2013/MIT-2850 NANOFRONTMAG and by the European Commission AMPHIBIAN (H2020-NMBP-2016-720853).

**References**

1 M. I. Dyakonov, and V. I. Perel, *JETP Lett.* **13** 467 (1971).
2 J. E. Hirsch, *Phys. Rev. Lett.* **83**, 1834 (1999).
3 R. Karplus, and J. M. Luttinger, Phys. Rev. **95**, 1154 (1954).
4 J. Sinova, D. Culcer, Q. Niu, N. A. Sinitsyn, T. Jungwirth, and A. H. MacDonald, *Phys. Rev. Lett.* **92**, 126603 (2004).
5 J. Sinova, S. O. Valenzuela, J. Wunderlich, C. H. Back and T. Jungwirth, *Rev. Mod. Phys.* **87**, 1213 (2015).
6 M. Liniers, V. Madurga, M. Vázquez and A. Hernando, *Phys. Rev. B* **31**, 4425 (1985).
7 S. O. Valenzuela, M. Tinkham, *Nature* **442**, 176 (2006).
8 Y. Kato, R. C. Myers, A. C. Gossard, D. D. Awschalom, *Science* **306** 1910 (2004).
9 K. Ando, and E. Saitoh, *Nature Comm.* **3**, 629 (2012).
10 H. L. Wang, C. H. Du, Y. Pu, R. Adur, P. C. Hammel, and F. Y. Yang *Phys. Rev. Lett.* **112**, 197201 (2014).
11 A. Hoffmann, *IEEE Trans. On Magnetics*, **49**, 5172 (2013).
12 N.F. Mott *Proc. Roy. Soc. A* **124** 425 (1929).
13 J. Smit, *Physica* **21**, 877 (1955).
14 J. Smit, *Physica* **24**, 39 (1958).
15 N. A. Sinitsyn, *Journal of Phys.: Cond. Matt.* **20**, 023201 (2008).
16 J. Ionue, G. E. W. Bauer, and L. W. Molenkamp, *Phys. Rev. B* **70**, 041303 (2004).
17 H. Kontani, T. Tanaka, D. S. Hirashima, K. Yamada, and J. Inoue, *Phys. Rev. Lett.* **102**, 016601 (2009).
18 E. Saitoh, M. Ueda, H. Miyajima, G. Tatara. *Appl. Phys. Lett.* **88** 182509 (2006).

**Supplementary Information.**

*Calculation of the magnetic field in a thin slab.* We analyze a slab which is infinite in the $y$ direction, and with dimensions $\ell_x, \ell_z$ in the $x, z$ directions. Then, the field at a position $(x, y, z)$ inside the slab, $-\ell_x/2 \leq x \leq \ell_x/2$, $-\ell_z/2 \leq z \leq \ell_z/2$ can be written, using the Biot Savart law, as

$$\vec{B}(x,y,z) = \frac{j_y}{c} \int_{-\infty}^{\infty} dy' \int_{-\ell_x/2}^{\ell_x/2} dx' \int_{-\ell_z/2}^{\ell_z/2} dz' \frac{(z-z')\hat{x} - (x-x')\hat{z}}{[(x-x')^2 + (y-y')^2 + (z-z')^2]^{3/2}} \quad (A1)$$

Where $\hat{x}$ and $\hat{z}$ are unit vectors along the $x$ and $z$ directions. Integrating over $x'$ and $y'$, we obtain

$$B_x(x,y,z) = \frac{2j_y}{c} \int_{-\ell_z/2}^{-\ell_z/2} dz' \tan\left(\frac{x-x'}{z-z'}\right)\Big|_{x'=-\ell_x/2}^{x'=\ell_x/2} \approx \frac{2\pi j_y}{c} \int_{-\ell_z/2}^{-\ell_z/2} dz' \text{sign}(z-z') \approx \pm \frac{2\pi j_y \ell_z}{c} \quad (A2)$$

$$B_z(x,y,z) = -\frac{2j_y}{c} \int_{-\ell_z/2}^{-\ell_z/2} dz' \frac{1}{2}\log[(x-x')^2 + (z-z')^2]\Big|_{x'=-\ell_x/2}^{x'=\ell_x/2} \approx \frac{2\pi j_y \ell_z}{c} \log\left|\frac{x+\ell_x/2}{x-\ell_x/2}\right| \quad (A3)$$

Where, in order to obtain the last result, the assumption $\ell_z \ll \ell_x$ has been made. The two signs in the expression for $B_x$ refer to the top and bottom surfaces of the slab. Note that the integral of the field along a contour defined as a section of the stripe of width $2x$ is given by

$$\Phi(x) \approx 2\int_{-x}^{x} dx\, B_x(x,y,z) \approx \frac{8\pi j_y x \ell_z}{c} \qquad (A4)$$

*Spin Hall angle due to the current induced magnetic field.* The spin Hall angle can be written as

$$\alpha \approx \frac{F_I}{eE_\parallel} \qquad (A5)$$

Where $F_I$ is given in eq.(4) of the main text, and $E_\parallel$ is the field which induces the longitudinal current

$$j_\parallel = nv \approx \frac{neE_\parallel \tau}{m} = \frac{neE_\parallel \ell}{mv_F} \qquad (A6)$$

Where $n$ is the carrier density, $v$ is the drift velocity, $\tau$ is the scattering time, $v_F$ is the Fermi velocity, and $\ell$ is the mean free path. For a three dimensional metal and using a nearly free electron description, we have

$$n = \frac{k_F^3}{3\pi^2}$$

$$v_F = \frac{\hbar k_F}{m} \qquad (A7)$$

Where $k_F$ is the Fermi wavevector. From eqs.(A5), (A6), and (A7) we obtain

$$\alpha \approx \frac{8e^2}{mc^2}\frac{k_F^2}{3\pi^2}\frac{\ell \ell_z}{\ell_x} \approx \frac{8}{137^2}\frac{k_F a_B}{3\pi^2} k_F \ell \frac{\ell_z}{\ell_x} \qquad (A8)$$

Where $a_B \approx 0.53$ Å is the Bohr radius.

For a two dimensional system where only one subband is occupied, the carrier density is

$$n = \frac{k_F^2}{2\pi \ell_z} \qquad (A9)$$

The spin Hall angle becomes

$$\alpha_{2D} \approx \frac{8}{137^2}\frac{1}{2\pi} k_F \ell \frac{a_B}{\ell_x} \qquad (A10)$$

*Skew scattering and mean free path.* The contribution from skew scattering to the spin Hall conductivity arises from the spin dependence of the scattering rates in the presence of the spin-orbit interaction[3,4,5]. We use a nearly free electron approximation to the electron band, and assume that the impurity induced matrix elements between scattering states can be written as

$$\langle \vec{k}', s'|V|\vec{k}, s\rangle = V_{k,k'}\left[\delta_{s,s'} + \frac{i\hbar^2}{4m^2c^2}\left(\langle s'|\vec{\sigma}|s\rangle \times \vec{k}'\right)\cdot \vec{k}\right] \qquad (A11)$$

We analyze a quasi-two dimensional system, where the Fermi surface can be approximated by a circle of radius $k_F$. The momenta $\vec{k}$ and $\vec{k}'$ lie in the $x-y$ plane. The transition rates near the Fermi surface can be written as function of the angle $\theta$ between $\vec{k}$ and $\vec{k}'$. Using eq.(A11), and considering higher order scattering processes, we approximate the transition rates by

$$W_\theta^{s_z} \approx \frac{1}{\tau} + \frac{1}{\tau_{SOI}}\frac{\hbar^2 k_F^2 s_z}{m^2 c^2}\sin(\theta) \qquad (A12)$$

where $s_z = \pm\frac{1}{2}$ describes the spin of the electron, and we have distinguished two scattering processes described by two scattering rates, $\tau$, which describe symmetric scattering, and $\tau_{SOI}$ which describes skew scattering induced by impurities with spin-orbit coupling. The transport properties can be computed from the Boltzmann equation, which can be written as

$$\frac{\partial f_\theta^{s_z}}{\partial t} = 0 = e\vec{E}\vec{v}\frac{\partial f_0(\epsilon)}{\partial \epsilon} + \sum_{\theta'} W_{\theta-\theta'}^{s_z}\left(f_\theta^{s_z} - f_{\theta'}^{s_z}\right) \quad (A13)$$

where $f_0(\epsilon)$ is the unperturbed Fermi-Dirac distribution and $\vec{v} = \frac{\hbar\vec{k}}{m}$. There are two independent equations which describe the distribution of electrons with $s_z = \pm\frac{1}{2}$. We split $f_\theta$ into a $s_z$ independent term, $\bar{f}_\theta$, and a term which describes the skew scattering, $\pm\delta f_\theta$, where the sign refers to the spin. We assume that $|\bar{f}_\theta| \gg |\delta f_\theta|$ and obtain

$$\bar{f}_\theta \approx -\frac{\hbar k_F e |\vec{E}| \tau \cos(\theta)}{m}\frac{\partial f_0(\epsilon)}{\partial \epsilon} \quad (A14)$$

The shift in the carriers' velocity is

$$v_\parallel \approx \frac{e|\vec{E}|\tau}{m} \quad (A15)$$

Inserting this expression into the Boltzmann equation, we obtain

$$0 = \frac{\hbar k_F e |\vec{E}| \cos(\theta)}{m}\frac{\partial f_0(\epsilon)}{\partial \epsilon} + \frac{1}{\tau}(\bar{f}_\theta \pm \delta f_\theta) \pm \frac{\hbar^2 k_F^2}{2m^2c^2}\frac{\tau}{\tau_{SOI}}\frac{\hbar k_F e|\vec{E}|\sin(\theta)}{m}\frac{\partial f_0(\epsilon)}{\partial \epsilon} \quad (A16)$$

This equation can be understood as if the function $\bar{f}_\theta \pm \delta f_\theta$ is induced by a field which is rotated with respect to the applied field by an angle

$$\alpha \approx \pm\frac{\hbar^2 k_F^2}{2m^2c^2}\frac{\tau}{\tau_{SOI}} \approx \pm\frac{(k_F a_B)^2}{137^2}\frac{\ell}{\ell_{SOI}} \quad (A17)$$

Where $a_B$ is the Bohr radius, and $\ell_{SOI}$ is a mean free path related to the impurities which induce the skew scattering.

Note that the estimate in eq.(A17) is valid only when one two dimensional subband is occupied. In the limit $\ell_z^{-1} \ll k_F$ the sample must be approximated using a three dimensional Fermi surface. In that case, the ansatz for the spin-orbit matrix elements defined in eq.(A5) does not favor a particular spin direction, when scattering processes are averaged over the Fermi surface. As a result, an unpolarized current will not lead to a net spin polarization at the edges of the sample.

*Effect of Lorentz force due to the field produced by the primary current $j_x$.* The field associated with the current $B_z(x)$ exerts a force on the carriers that move along the $x$ axis. This force is given by $F_x(x) = evB_z(x)/c$ and induces a perturbation of the uniform density of carriers. However the value of the maximum Lorentz force for the density currents considered in this article (Hall effect with an applied field of ~1 G) is four orders of magnitude smaller than $F_x(x) = \mu_B \partial_x B_z(x)$. Therefore, the charge imbalance generated by the magnetic field induced by the current can be neglected.

*Experimental techniques capable of measuring the contribution of the magnetic field induced by the current to the spin Hall effect.*
Ref.(2) in the main text argues that the magnetic field induced by a current will not generate a voltage on a metallic bridge which contacts both edges of a Hall bar. This argument is correct, and it is based on the fact that, when the bridge is close to the Hall bar, the magnetic field induced by the current will thread both the bar and the bridge, preventing any spin current to flow along the bar, and to induce a voltage across it. On the other hand, the combination of a spin gradient and spin dependent scattering by impurities will lead to the voltage discussed in ref.(2).

This situation changes if the magnetic field threading the bridge is different from the field threading the bar. This can be achieved by changing the distance between the bridge and the Hall bar. Any technique which does not require a contact between the edges of the bar will be sensitive to the spin accumulation induced by the field associated to the current. Possible experimental techniques are. I) injection of a spin polarized current using a ferromagnet, refs.(7), and (9) of the main text, ii) spin pumping from a magnetic insulator, ref.(10) of the main text, iii) measurement of the Kerr rotation by optical means, ref. (8) of the main text. A general overview of techniques used to measure the spin Hall effect can be found in the reviews cited in the main text, refs.(5) and (11).